\documentclass[a4paper,twocolumn,english,aps]{revtex4}
\usepackage{graphicx}
\usepackage{amssymb}

\makeatletter

\usepackage{babel}
\makeatother
\begin{document}

\def\eqd{\,{\buildrel d \over =}\,}\def\subst{\,{\buildrel {y'=y f(\tau^{-1})} \over =}\,}\textsf{}

\title{A method for extracting the scaling exponents of a self-affine, non-Gaussian
process from a finite length timeseries.}

\author{K. Kiyani}

\email{k.kiyani@warwick.ac.uk}

\author{S. C. Chapman}

\author{B. Hnat}

\affiliation{Centre for Fusion, Space and Astrophysics; Dept. of Physics, University
of Warwick, Coventry CV4 7AL, UK}

\date{25 July 2006}

\begin{abstract}
We address the generic problem of extracting the scaling exponents
of a stationary, self-affine process realised by a timeseries of finite
length, where information about the process is not known \emph{a priori}.
Estimating the scaling exponents relies upon estimating the moments,
or more typically structure functions, of the probability density
of the differenced timeseries. If the probability density is heavy
tailed, outliers strongly influence the scaling behaviour of the moments.
From an operational point of view, we wish to recover the scaling
exponents of the underlying process by excluding a minimal population
of these outliers. We test these ideas on a synthetically generated
symmetric $\alpha$-stable L\'{e}vy process and show that the L\'{e}vy exponent
is recovered in up to the $6^{th}$ order moment after only $\sim$0.1-0.5\%
of the data are excluded. The scaling properties of the excluded outliers
can then be tested to provide additional information about the system.
\end{abstract}
\maketitle

\section{introduction}

There is increasing observational evidence that natural systems often
show scaling in a statistical sense, coincident with non-Gaussian
`heavy tailed' statistics. Complex systems approaches aim to understand
these phenomena as universal, with a key quantitative prediction of
theory being scaling exponents. Importantly, the identification of
universal scaling functions implies the ability to describe many different
length and time scales as well as apparently disjoint physical phenomena
with the same macroscopic scaling behaviour \cite{Sethna2001,Sornette2000,mandelbrot1983}.

One of the outstanding challenges in complex system science is then
to find robust methods that (i) establish whether there is scaling
and (ii) accurately determine the scaling exponents for statistical
measures of series of data that are of large, but finite length. We
seek to determine the scaling properties of probability distributions
that are heavy-tailed. The scaling exponents can be determined through
the scaling behaviour of the moments, usually characterised by computing
structure functions. Where the probability density is heavy tailed
the moments and structure functions can depend strongly on extremal
values, or outliers. Once we insist that the data series is represented
by a finite number of measurements, the values at which these outliers
occur will always vary between one realisation and the next. From
an operational point of view, that is, when the underlying behaviour
is not known \emph{a priori}, these outliers can potentially distort
the scaling properties of the data and the values of scaling exponents
extracted via the structure functions. In this paper we propose a
generic method for excluding these outliers in a manner which does
not distort the underlying scaling properties of the data. These outliers
also contain information and we explore a method for extracting this.
We will test these ideas on numerically generated L\'{e}vy processes.

There has been considerable interest in fractional kinetics as providing
stochastic models for the data of candidate complex systems \cite{zaslavsky2002,SchmittSchertzer1999}.
L\'{e}vy processes have been identified for example in biological systems
(foraging of albatrosses \cite{Viswanathan2002}), financial markets
(S\&P 500 \cite{MantegnaStanley1995}) and physical systems (laser
cooling and trapping \cite{BardouBouchaud02}). A robust method for
determining the L\'{e}vy exponent from finite sized data sets, where the
statistics are not known \emph{a priori} is thus important in its
own right. The method that we propose here is however quite generic,
with application to a wide class of systems that show scaling; for
example those that can be modelled by stochastic differential equations
with scaling \cite{Hnat2003,Hnat2005,Chapman2005}. In this wider
context L\'{e}vy processes, which have non-convergent higher order moments,
provide a particularly stringent test of our ideas.

\subsection{Statistical self-similarity}

One can characterise fluctuations in a timeseries $x(t)$ on a given
time scale $\tau$ in terms of a differenced variable $y(t,\tau)$
\begin{equation}
y(t,\tau)=x(t+\tau)-x(t)\ ,\label{eq:eqn1}
\end{equation}
for time $t$ and interval $\tau$, where the timeseries/stochastic
process $x(t)$ represents a particular realisation or set of observations
of the system from which the $y$'s are generated. We consider the
case where the $y(t,\tau)$ satisfy the following scaling relation
\begin{equation}
y(b\tau)\eqd f(b)y(\tau)\ ,\label{eq:eqn2}
\end{equation}
 where $b$ is some scale dilation factor; $\eqd$ indicates an equality
in the statistical/distribution sense; $f$ is some scaling function
(to be determined); and we have dropped the time argument in the increments
$y$ by assuming statistical stationarity. Both $b$ and $f(b)$ are
positive. The property in (\ref{eq:eqn2}) is a generalized form of
\emph{self-affinity,} and in this sense $x(t)$ is a \emph{self-affine}
field. Self-affinity is a particular case of statistical self-similarity
i.e. stochastic processes that exhibit the absence of characteristic
scales \cite{mandelbrot1983,GreisGreenside1991,Chapman2005}. We can
write the scaling transformations (\ref{eq:eqn2}) as
\begin{equation}
\tau'=b\tau\ ,\quad\; y'=f(b)y\ ,\label{eq:eqn3}
\end{equation}
where the primed variables represent scaled quantities. Conservation
of probability under change of variables implies that the probability
density function (PDF) of $y$, $P(y,\tau)$ is related to the PDF
of $y'$, $P(y',\tau')$ by
\begin{eqnarray}
P(y,\tau) & = & P'(y',\tau')\left|\frac{dy'}{dy}\right|\ ,\label{eq:eqn4}
\end{eqnarray}
thus giving from (\ref{eq:eqn3})
\begin{equation}
P(y,\tau)=f(b)P'(f(b)y,\tau')\ .\label{eq:eqn4pt5}
\end{equation}
 The result (\ref{eq:eqn4pt5}) expresses the fact that the stochastic
process $x(t)$ is statistically self-similar i.e. that a given process
on scale $\tau'$ (and thus $y'$) maps onto another process based
on a different scale $\tau$ (and $y$) by the scaling transformation
in (\ref{eq:eqn3}); and that the PDFs of both these processes are
related by (\ref{eq:eqn4pt5}).

We can go further and reduce the expression (\ref{eq:eqn4pt5}) to
a function of one variable. Since the dilation factor $b$ is arbitrary
we choose $b=\tau^{-1}$, which gives the important result
\begin{eqnarray}
P(y,\tau) & = & f(\tau^{-1})P'(f(\tau^{-1})y,1)\nonumber \\
 & = & f(\tau^{-1})\mathcal{P}_{s}(f(\tau^{-1})y)\ ,\label{eq:eqn5}
\end{eqnarray}
and shows that any PDF $P$ of increments $y$ characterised by a
time increment $\tau$ may be collapsed onto a single unique PDF $\mathcal{P}_{s}$
of rescaled increments $f(\tau^{-1})y$ and time increment $\tau=1$,
by the above scaling transformation. Identification of this unique
\emph{scaling function} and the ensuing collapse is a clearer method
of discriminating between different (universality) scaling classes
than simply identifying the scaling exponents by themselves \cite{Sethna2001}.

In this paper we will consider the scaling as defined by the structure
functions. The generalised structure functions of order $p$ are simply
defined as
\begin{equation}
S^{p}(\tau;\pm\infty)=\left\langle \left|y\right|^{p}\right\rangle 
\ =\int_{-\infty}^{\infty}\left|y\right|^{p}P(y,\tau)dy\ .
\label{eq:eqn5pt5}
\end{equation}
The analysis which follows is also valid for the moments; however,
structure functions are typically calculated for data. This avoids
the result that odd order moments of symmetric PDFs are zero so that
as a consequence, in a physical system, they would be dominated by
experimental error. Using the transformation (\ref{eq:eqn5}), the
scaling of the structure functions is:
\begin{eqnarray}
S^{p}(\tau;\pm\infty) & = & 
\int_{-\infty}^{\infty}\left|y\right|^{p}P(y,\tau)dy\nonumber \\
 & = & \int_{-\infty}^{\infty}\left|y\right|^{p}
f(\tau^{-1})\mathcal{P}_{s}(f(\tau^{-1})y)\  dy\nonumber \\
 & \subst & \left(f(\tau^{-1})\right)^{-p}
\int_{-\infty}^{\infty}\left|y'\right|^{p}\mathcal{P}_{s}(y')\  dy'\nonumber \\
 & = & \left(f(\tau^{-1})\right)^{-p}\mathcal{S}_{s}^{p}(1;\pm\infty)\ .
\label{eq:eqn5pt75}
\end{eqnarray}
This formalism encompasses a general class of self-affine systems
in the sense that it is not restricted to the well-studied case of
mono-exponent scaling. 

The above result (\ref{eq:eqn5pt75}) holds provided that the PDF
$P$ is defined for all $y$. However, for finite data sets this is
not the case. In this situation we have the integral (\ref{eq:eqn5pt5})
defined for the interval $[y_{-},y_{+}]$ where the $y_{\pm}$ are
defined in some sense by the largest events measured in the data set.
The values of $y_{\pm}$ will depend on the time scale $\tau$ and
the sample size $N$ (which will be held constant). Thus the structure
functions for the finite data set are
\begin{equation}
S^{p}(\tau;y_{\pm}(\tau))=\int_{y_{-}(\tau)}^{y_{+}(\tau)}\left|y\right|^{p}P(y,\tau)dy\ .
\label{eq:eqn5pt9}
\end{equation}
Manipulating this in a similar way to (\ref{eq:eqn5pt75}) results
in the following scaling relation
\begin{equation}
S^{p}(\tau;y_{\pm}(\tau))=\left(f(\tau^{-1})\right)^{-p}
\mathcal{S}_{s}^{p}(1;y_{\pm}(\tau)f(\tau^{-1}))\ .
\label{eq:eqn5pt95}
\end{equation}
If we assume that the values $y_{\pm}$ scale with $\tau$ in the
same way as the increments $y$ in (\ref{eq:eqn3}), then (\ref{eq:eqn5pt95})
becomes:
\begin{equation}
S^{p}(\tau;y_{\pm}(\tau))=\left(f(\tau^{-1})\right)^{-p}\mathcal{S}_{s}^{p}(1;y_{s\pm}(1))\ .
\label{eq:eqn5pt97}
\end{equation}

We will consider the case of self-affine scaling where the scaling
function $f$ takes the form of a mono-scaling power law $f(b)=b^{H}=\tau^{-H}$,
where $H$ is known as the \emph{Hurst} exponent. Equation (\ref{eq:eqn5})
then becomes
\begin{equation}
P(y,\tau)=\tau^{-H}\mathcal{P}_{s}(\tau^{-H}y)\ ,
\label{eq:5pt98}
\end{equation}
and (\ref{eq:eqn5pt75}) becomes
\begin{equation}
S^{p}(\tau;\pm\infty)=\tau^{\zeta(p)}\mathcal{S}_{s}^{p}(1;\pm\infty)\ ,
\label{eq:eqn5pt85}
\end{equation}
where $\zeta(p)=Hp$ for this self-affine case. A log-log plot of
$S^{p}$ vs. $\tau$ for various orders $p$ reveals scaling if present,
and the slope of such a plot determines the exponents $\zeta(p)$
\cite{Sornette2000,BohrJensen1998}. One then verifies that $\zeta(p)=Hp$
by plotting $\zeta(p)$ as a function of $p$. 

The aim of this paper is to obtain a good estimate of the scaling
properties of (\ref{eq:eqn5pt5}), the structure functions at $N\rightarrow\infty$,
via (\ref{eq:eqn5pt97}) for $N$ large but finite. However, we can
anticipate that simply setting the limits $y_{\pm}$ of the integral
(\ref{eq:eqn5pt9}) to the largest values found in a given realisation
of the data, will give a scaling behaviour of (\ref{eq:eqn5pt97})
which can differ substantially from that of (\ref{eq:eqn5pt85}).
This problem arises since the $y$ values of the extremal points fluctuate
between one realisation and the next, and these fluctuations are more
significant in heavy tailed distributions. This in turn will strongly
modify the integral. We will therefore explore the possibility of
choosing a range for the integral (\ref{eq:eqn5pt9}) based on the
scaling property of the data itself, by systematically excluding the
most extreme outlying points. This has the added advantage of not
requiring \emph{a priori} information about the system.

We stress that as our aim is to extract scaling exponents, we do not
attempt to estimate the value of the moments or structure functions.
Thus we will not compute an estimate of the integral (\ref{eq:eqn5pt5})
\emph{per se}, rather we will examine methods for quantifying its
dependence on the dilation factor $b$ (or equivalently $\tau$).
Hence, our method can be applied to L\'{e}vy processes -- where the moments
are not defined, but where the PDF has scaling.

The paper is organised as follows. We first introduce the L\'{e}vy process
that we will use to obtain (\ref{eq:eqn5pt9}) and briefly survey
results pertaining to its asymptotic behaviour. We then discuss the
effects of finite sized data sets and demonstrate the effect of removing
outliers on the scaling behaviour of the L\'{e}vy process. We then explore
the behaviour of these outliers.

\section{L\'{e}vy processes and Finite Size Effects}

\subsection{$\alpha$- stable processes}

Many stochastic processes exhibit self-affine scaling and are characterised
by `broad tails' described by power-laws in their PDFs. Some possible
mechanisms by which these power laws occur are discussed in \cite{Sornette2000}.
This general class of stochastic processes can be described in the
context of so-called \emph{$\alpha$-stable L\'{e}vy processes} 
\cite{samorodnitsky1994,zaslavsky2002,JanickiWeron1994}.
We will restrict our attention to symmetric $\alpha$-stable processes.
The PDFs $L_{\alpha}^{\gamma}$ of the increments $y$ of these processes
are defined through the Fourier transform of their characteristic
function 
\begin{equation}
L_{\alpha}^{\gamma}(y,\tau)=\frac{1}{2\pi}
\int_{-\infty}^{\infty}dke^{iky}e^{-\gamma\tau|k|^{\alpha}},
\label{eq:eqn6}
\end{equation}
where $\gamma\geq0$ and $\tau\geq0$ are the characteristic scales
of the process and describe the width of the distribution; and $\alpha\in(0,2]$
parameterises the stability of the distribution; $\alpha$ can be
heuristically seen as an indication of the variability of the increments
of such processes (also known as \emph{L\'{e}vy flights}). In this paper
we will take $\gamma=1$ and will consequently reduce the notation
$L_{\alpha}^{\gamma}$ to $L_{\alpha}$. The form and convention of
the parameters in equation (\ref{eq:eqn6}) are similar to that presented
in \cite{PaulBaschnagel1999}; for a more rigorous discussion of the
mathematical properties of such processes readers are referred to
\cite{samorodnitsky1994,JanickiWeron1994}. 

\begin{figure}
\begin{center}a)\includegraphics[%
  width=1.0\columnwidth,
  keepaspectratio]{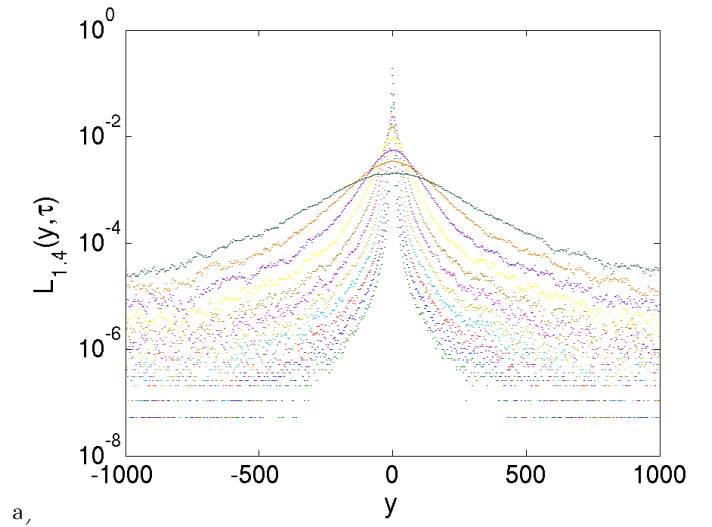}
\end{center}
\begin{center}b)\includegraphics[%
  width=1.0\columnwidth,
  keepaspectratio]{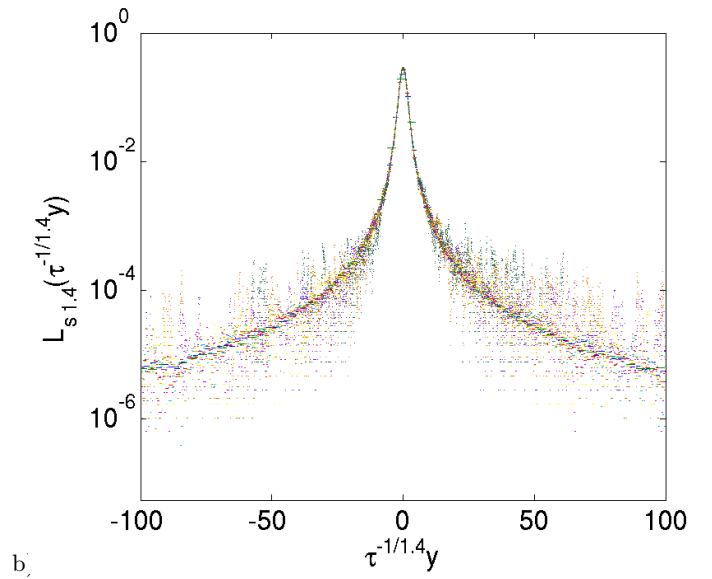}
\end{center}
\caption{\label{cap:LevyCollapse}Plots showing probability density functions
of the L\'{e}vy distribution for index $\alpha=1.4$ ($N=10^{6}$) at
different values of differenced interval $\tau$ (a) before and (b)
after the scaling collapse described by (\ref{eq:eqn7}). }
\end{figure}

From (\ref{eq:eqn6}) it follows that the scaling properties of $L_{\alpha}$
are
\begin{eqnarray}
L_{\alpha}(y,\tau) & = & \tau^{-\frac{1}{\alpha}}
L_{\alpha}(\tau^{-\frac{1}{\alpha}}y,1)\nonumber \\
 & = & \tau^{-\frac{1}{\alpha}}\mathcal{L}_{s,\alpha}(\tau^{-\frac{1}{\alpha}}y)\ ,
\label{eq:eqn7}
\end{eqnarray}
from which the Hurst exponent of symmetric $\alpha$-stable processes
is $H=1/\alpha$, by comparison with (\ref{eq:5pt98}). Figure \ref{cap:LevyCollapse}
(a) shows the $L_{\alpha}(y,\tau)$ for $\alpha=1.4$ and a range
of $\tau=2^{0},2^{1},\ldots,2^{10}$; the scaling collapse (\ref{eq:eqn7})
has been applied to these in Figure \ref{cap:LevyCollapse} (b).

We now focus on the asymptotic behaviour of such distributions. By
expanding the complex exponential in equation (\ref{eq:eqn6}) and
integrating one can show that in the large $y$ limit we obtain the
asymptotic behaviour
\begin{eqnarray}
\lim_{y\rightarrow\infty}L_{\alpha}(y,\tau) & \simeq & 
\frac{\tau\Gamma(1+\alpha)\sin(\pi\alpha/2)}{\pi\left|y\right|^{1+\alpha}}\nonumber \\
 & = & D_{\alpha}\frac{\tau}{\left|y\right|^{1+\alpha}}\ .
\label{eq:eqn8}
\end{eqnarray}
for $y\gg\tau^{\frac{1}{\alpha}}$ \cite{PaulBaschnagel1999,Jespersen1999}.
It immediately follows that these power-law tails ensure that for
the $p^{th}$ moment to exist, $p-\alpha<0$. Hence the process has
no variance defined for $0<\alpha<2$, and in the cases where $0<\alpha\leq1$
the process will also have no mean defined i.e. both these quantities
and the other higher order moments are infinite.

A generalized version of the Central Limit Theorem (CLT) \cite{Sornette2000}
ensures that the sum of all independent and identically distributed
(i.i.d.) random variables with no finite variance that have distributions
with power law tails that go asymptotically as $y^{-1-\alpha}$ ($\alpha\in(0,2]$),
will converge to a L\'{e}vy distribution of the same index $\alpha$.
In practice, however, we will always obtain a finite mean and variance
from a finite length timeseries.

\subsection{Finite-Size effects and outliers}

We will now consider in detail the procedure for extracting the scaling
exponents, $\zeta(p)$, from the structure functions in (\ref{eq:eqn5pt85}).
This centres on first computing $S^{p}(\tau;y_{\pm})$ and the gradients
$\zeta(p)$ of log-log plots of $S^{p}(\tau;y_{\pm})$ vs. $\tau$.
If the process is self-affine ($\zeta(p)=Hp$) we should obtain a
straight line on a plot of $\zeta(p)$ vs. $p$ from which we can
measure the gradient and obtain the Hurst exponent, $H$. Note that
the $\zeta(p)$ for several $p$ are needed to determine $H$ uniquely
\cite{Chapman2005}. 

However, finite sample sizes result in pseudo multi-affine behaviour.
As we will show, the primary reason for this anomalous behaviour is
due to the large scatter in the outlying events of the tails of the
distribution. In the case of L\'{e}vy-like processes this scaling bias
shows up as a saturation/roll-over on the $\zeta(p)$ plots at $p>\alpha$.
\begin{figure}
\begin{center}a)\includegraphics[%
  width=1.0\columnwidth,
  keepaspectratio]{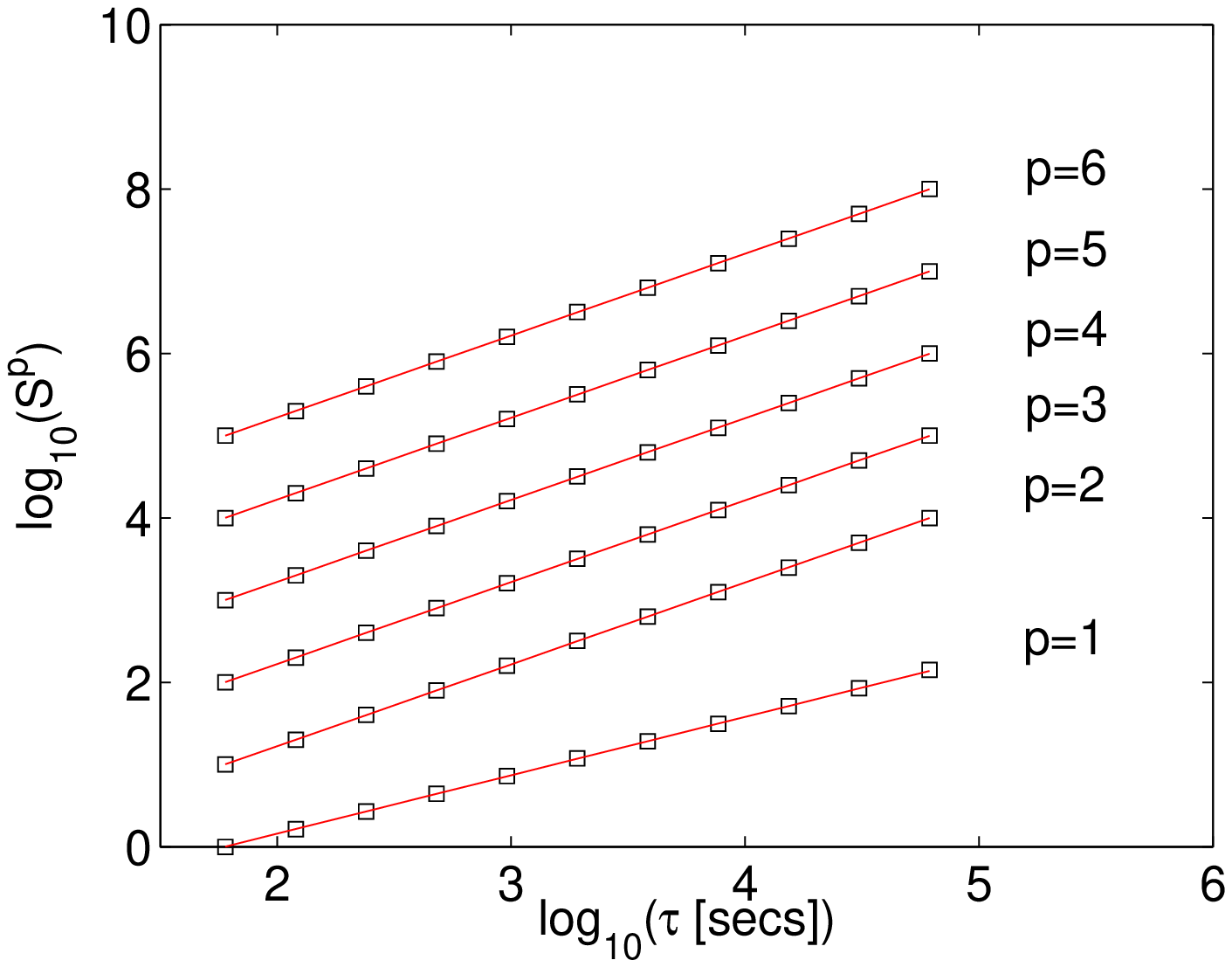}
\end{center}
\begin{center}b)\includegraphics[%
  width=1.0\columnwidth,
  keepaspectratio]{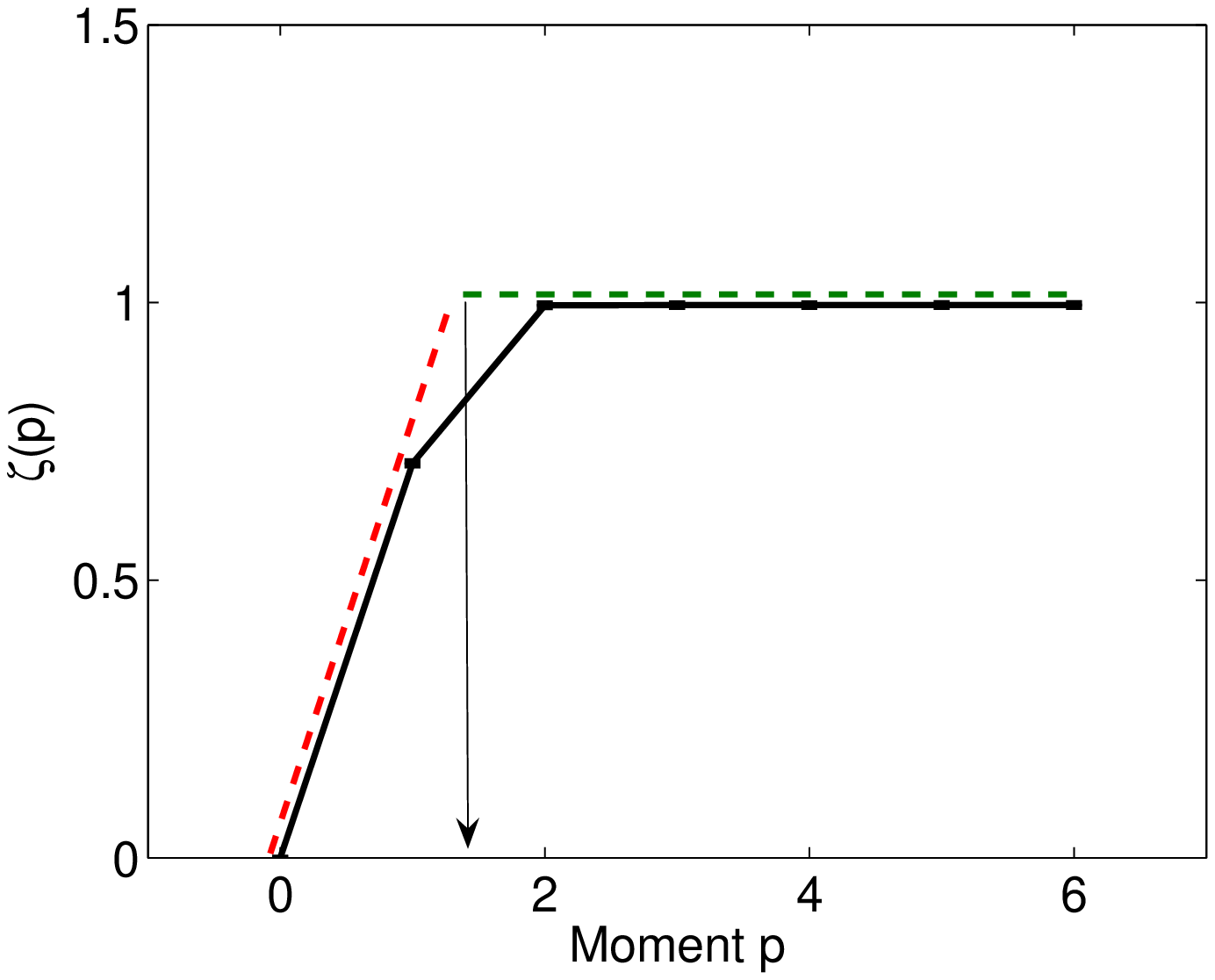}
\end{center}
\caption{\label{cap:levyRollOver}Plots of (a) generalised structure functions
$S^{p}$ vs. $\tau$ for moments of order $p=1-6$, and (b) the scaling
exponents $\zeta(p)$ vs. $p$ (solid black line). These quantities
are shown for a L\'{e}vy process of index $\alpha=1.4$ and with $N=10^{6}$
data points. The dashed red line indicates the expected scaling $\zeta(p)=p/\alpha$
for $p<\alpha$; the green dashed line indicates the scaling exponent
observed for $p>\alpha$ in a finite sized sample. The vertical arrow
at $p\simeq\alpha$ seperates these two regions of scaling.}
\end{figure}
This can be seen in Figure \ref{cap:levyRollOver} which illustrates
both the methodology of extracting scaling exponents from structure
function plots, and this finite sample size saturation effect in a
L\'{e}vy process of index $\alpha=1.4$. This saturation effect is well-known
and an explanation for it can be found in the work by Schmitt \emph{et.
al.} \cite{SchmittSchertzer1999} and Chechkin and Gonchar \cite{ChechkinGonchar2000}.
We will now establish the scaling properties of these extremal events.
We need to emphasise, however, that in contrast to \cite{SchmittSchertzer1999,ChechkinGonchar2000}
we will propose a method for estimating the integral in (\ref{eq:eqn5pt5})
such that the scaling in (\ref{eq:eqn5pt85}) is recovered for \emph{all}
$p$.

We consider the situation where we have many realisations, that is
many data series of size $N$ obtained from the same process. Each
of these realisations will have extremal points $y^{*}$ of their
respective PDF. We know the properties of $\bar{y}^{*}$, the ensemble
average of the $y^{*}$ over the realisations, since it will fall
on the L\'{e}vy asymptotic distribution (\ref{eq:eqn8}). We will use
a simple example of Extreme Value Theory, EVT, (see \cite{Sornette2000})
to obtain an estimate of the largest event in a sample of $N$ i.i.d.
measurements of a random variable $y\in\mathbb{R}^{+}$. An approximation
to the probability to see an event that occurs only once can be made
by realising that an event with probability $P$ occurs typically
$NP$ times. Therefore, the rarest event in a sample of $N$ measurements,
which occurs typically only once can be seen to be described by $NP(y\geq\bar{y}^{*})=1$,
where $P(y\geq\bar{y}^{*})$ is the probability of observing an event
greater than or equal to $\bar{y}^{*}$ ; thus
\begin{equation}
P(y\geq\bar{y}^{*})=\frac{1}{N}\ .\label{eq:eqn9}
\end{equation}
We can generalise this to the $m^{th}$ largest event:
\begin{equation}
P(y\geq\bar{y}_{m}^{*})=\frac{m}{N}\ .\label{eq:eqn10}
\end{equation}
For the case of the L\'{e}vy-like process, within the limits of the integral
in $P(y\geq\bar{y}_{m}^{*})$ the main contribution is from the tail
and thus we can use (\ref{eq:eqn8}) and estimate $P(y\geq\bar{y}_{m}^{*})$
to be
\begin{equation}
P(y\geq\bar{y}_{m}^{*})=\int_{\bar{y}_{m}^{*}}^{\infty}L_{\alpha}(y,\tau)dy\simeq D_{\alpha}\tau\int_{\bar{y}_{m}^{*}}^{\infty}\frac{dy}{\left|y\right|^{1+\alpha}}\ .
\label{eq:eqn11}
\end{equation}
Evaluating the integral and equating with (\ref{eq:eqn10}) gives
the following result for the scaling behaviour of the $m^{th}$ largest
event
\begin{equation}
\bar{y}_{m}^{*}=\left(\frac{D_{\alpha}N\tau}{m\alpha}\right)^{\frac{1}{\alpha}}\ .
\label{eq:eqn12}
\end{equation}

A more detailed account would be to attempt to specify approximately
the full PDF of the $m^{th}$ largest event amongst $N$ i.i.d. measurements.
Following Sornette \cite{Sornette2000} the cumulative distribution
function (CDF) $\Pi(y<\bar{y}_{m}^{*})$ of the maximum value is
\begin{equation}
\Pi(y<\bar{y}_{m}^{*})=\int_{-\infty}^{\bar{y}_{m}^{*}}p_{N}(y)dy\simeq 
e^{-\frac{N}{m}P(y\geq\bar{y}_{m}^{*})}\ ,
\label{eq:eqn13}
\end{equation}
where $p_{N}(y)$ is the PDF of the maximum value among $N$ observations,
and is obtained by differentiating equation (\ref{eq:eqn13}) to obtain
\begin{equation}
\frac{d\Pi(y<\bar{y}_{m}^{*})}{dy_{m}^{*}}=p_{N}(\bar{y}_{m}^{*})=
\frac{N}{m}L_{\alpha}(\bar{y}_{m}^{*},\tau)e^{-\frac{N}{m}P(y\geq\bar{y}_{m}^{*})}\ .
\label{eq:eqn14}
\end{equation}
By substituting (\ref{eq:eqn11}) in (\ref{eq:eqn13}) we obtain an
estimate of the $m^{th}$ largest value, $\bar{y}_{m,\Pi}^{*}$, that
will not be exceeded with probability $\Pi$. By setting the LHS of
(\ref{eq:eqn13}) to some probability $0<\Pi<1$, we obtain
\begin{equation}
\bar{y}_{m,\Pi}^{*}=\left(\frac{D_{\alpha}N\tau}{m\alpha\ln(1/\Pi)}\right)^{\frac{1}{\alpha}}\ .
\label{eq:eqn15}
\end{equation}
If one was to set $\Pi=1/2$ the value of $y_{m}^{*}$ would correspond
to the median value of the $m^{th}$ largest event. To obtain the
modal value of $\bar{y}_{m}^{*}$ , we optimise for the maximum by
differentiating (\ref{eq:eqn14}) and setting it to zero. This gives
us the following solution for the modal value of $\bar{y}_{m}^{*}$
\begin{equation}
\bar{y}_{m,mode}^{*}=\left(\frac{D_{\alpha}N\tau}{m(1+\alpha)}\right)^{\frac{1}{\alpha}}\ .
\label{eq:eqn16}
\end{equation}

By comparing these expressions one can see that although the approximation
of $\bar{y}_{m}^{*}$ becomes more refined, the scaling with $\tau$
is still that of (\ref{eq:eqn12}). Thus we will proceed using the
simplest expression (\ref{eq:eqn12}). In addition, we will be working
with a varying fraction $m/N$ rather than varying $m$ or $N$ separately.
Importantly, since we are concerned primarily with the scaling with
respect to $\tau$ we will write $\bar{y}_{m}^{*}$ more informatively
as $\bar{y}_{m}^{*}(\tau)$ and thus adding to our scaling relations
\begin{equation}
\bar{y}_{m}^{*}(\tau)=\tau^{\frac{1}{\alpha}}\bar{y}_{m}^{*}(1)\ ,
\label{eq:eqn17}
\end{equation}
as expected from equation (\ref{eq:eqn2}) 
\footnote{note that the distributional equality $\eqd$ is not needed here as
$\bar{y}_{m}^{*}$ is a statistical quantity.
}
. We emphasise that this is the scaling of $\bar{y}_{m}^{*}$; the
average over the $m^{th}$ largest events of a large number of realisations
(timeseries). In practice we will have a single realisation and thus
one value of $y_{m}^{*}$ which will fluctuate about this ensemble
averaged $\bar{y}_{m}^{*}$. The behaviour (\ref{eq:eqn17}) refers
to the property that any point in the curve $P(y,\tau)$ scales as
(\ref{eq:eqn5}) and (\ref{eq:eqn3}).

\section{Structure functions}

\subsection{Effects of finite sample size}

We can now investigate the scaling behaviour of the structure functions
of a L\'{e}vy-like process, but now with a finite sample size. Following
the procedure in (\ref{eq:eqn5pt97}) \textsf{}we can discuss the
structure functions in the average sense, that is averaged over many
realisations of our $N$ sample finite length timeseries: 
\begin{eqnarray}
\bar{S}^{p}(\tau;\bar{y}_{1,\pm}^{*}(\tau)) & = & 
\int_{-\bar{y}_{1,-}^{*}(\tau)}^{\bar{y}_{1,+}^{*}(\tau)}\left|y\right|^{p}
L_{\alpha}(y,\tau)dy\nonumber \\
 & = & \int_{-\bar{y}_{1,-}^{*}(\tau)}^{\bar{y}_{1,+}^{*}(\tau)}\left|y\right|^{p}
\tau^{-\frac{1}{\alpha}}\mathcal{L}_{s,\alpha}(\tau^{-\frac{1}{\alpha}}y)\  dy\nonumber \\
\label{eq:eqn18}
\end{eqnarray}
where we have set $m=1$ in $\bar{y}_{m}^{*}$ to emphasise that this
is the structure function for the raw data with the largest events
obviously bounding the data; the subscripts $+$ and $-$ indicate
the largest positive and negative events. The substitution $y'=\tau^{-\frac{1}{\alpha}}y$
gives
\begin{eqnarray}
\bar{S}^{p}(\tau;\bar{y}_{1,\pm}^{*}(\tau)) & = & \tau^{\frac{p}{\alpha}}
\int_{-\bar{y}_{1,-}^{*}(\tau)\tau^{-\frac{1}{\alpha}}}^{\bar{y}_{1,+}^{*}(\tau)
\tau^{-\frac{1}{\alpha}}}\left|y'\right|^{p}\mathcal{L}_{s,\alpha}(y')\  dy'\nonumber \\
 & = & \tau^{\frac{p}{\alpha}}\left[\int_{0}^{\bar{y}_{1,+}^{*}(\tau)
\tau^{-\frac{1}{\alpha}}}y'^{p}\mathcal{L}_{s,\alpha}(y')\  dy'\right.\nonumber \\
 &  & \ \left.+\int_{0}^{\bar{y}_{1,-}^{*}(\tau)
\tau^{-\frac{1}{\alpha}}}y'^{p}\mathcal{L}_{s,\alpha}(y')\  dy'\right]\ .
\label{eq:eqn19}
\end{eqnarray}
To approximate the integrals in (\ref{eq:eqn19}) we assume that values
of the largest events are deep in the tail region of the distribution
so that we may use the asymptotic form (\ref{eq:eqn8}). This gives
\begin{eqnarray}
\bar{S}^{p}(\tau;\bar{y}_{1,\pm}^{*}(\tau)) & = & 
\tau\ \frac{D_{\alpha}}{p-\alpha}\left(\bar{y}_{1,+}^{*(p-\alpha)}(\tau)
+\bar{y}_{1,-}^{*(p-\alpha)}(\tau)\ \right)\nonumber \\
 &  & \quad\quad\quad\quad\quad\quad\quad\quad\quad\quad\forall p>\alpha,
\label{eq:eqn20}
\end{eqnarray}
where the condition $p>\alpha$ is necessary as all structure functions
of order $p<\alpha$ of a L\'{e}vy distribution exist (i.e. are finite)
and this approximation would result in an infrared divergence in (\ref{eq:eqn19}),
which is clearly incompatible. For the ensemble average (\ref{eq:eqn11}),
(\ref{eq:eqn12}) and (\ref{eq:eqn17}) hold; thus we can simply substitute
(\ref{eq:eqn17}) into (\ref{eq:eqn20}) to obtain: 
\begin{equation}
\bar{S}^{p}(\tau;\bar{y}_{1,\pm}^{*}(\tau))=
\tau^{\frac{p}{\alpha}}\ \frac{D_{\alpha}}{p-\alpha}\left(\bar{y}_{1,+}^{*(p-\alpha)}(1)+
\bar{y}_{1,-}^{*(p-\alpha)}(1)\right)\ .
\label{eq:eqn21}
\end{equation}
 In practice the value of $y_{m}^{*}$ will vary for each realisation
of $P(y,\tau)$ about the average $\bar{y}_{m}^{*}$ which obeys (\ref{eq:eqn17}).
For a given functional form of $P(y,\tau)$ the $y_{m}^{*}$ will
have some probability density $p_{N}(y_{m}^{*})$ with a statistical
spread about the average $\bar{y}_{m}^{*}$. An approximation to this
can be made by substituting the asymptotic tail form of equation (\ref{eq:eqn8})
into equation (\ref{eq:eqn14}) to obtain
\begin{equation}
p_{N}(y_{m}^{*})=\frac{\Lambda}{y_{m}^{*1+\alpha}}
\exp\left(-\frac{\Lambda}{\alpha y_{m}^{*\alpha}}\right)\ ,
\label{eq:eqn22}
\end{equation}
where $\Lambda$ is given by
\begin{equation}
\Lambda=\frac{ND_{\alpha}\tau}{m}\ .
\label{eq:eqn23}
\end{equation}

\begin{figure}
\begin{center}\includegraphics[%
  width=1.0\columnwidth,
  keepaspectratio]{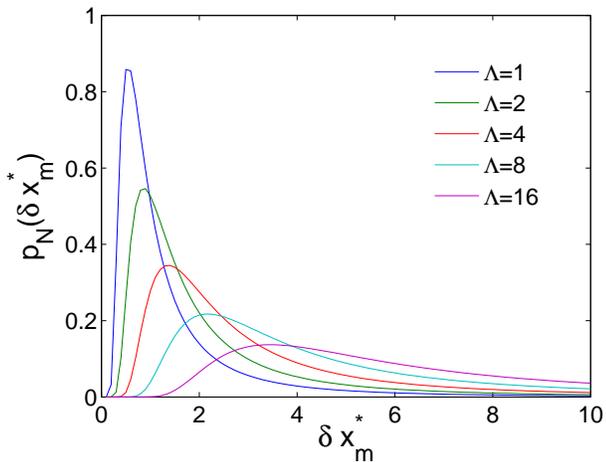}
\end{center}
\caption{\label{cap:evtPdf}Plot showing the PDF, in equation (\ref{eq:eqn22}),
of the $m^{th}$ largest value of a sample size $N$ of a set of measurements
taken from a L\'{e}vy-like process; the L\'{e}vy index $\alpha=1.5$.}
\end{figure}

Equation (\ref{eq:eqn22}) is of the form of a stretched exponential.
As with any power-law tailed PDF it has infinite variance for $0<\alpha<2$.
In the context of EVT, equation (\ref{eq:eqn22}) is not surprising
as it is simply an Extreme Value Distribution of Type II i.e. the
PDF from a Fr\'{e}chet distribution. The extreme value distributions can
be seen as the large event statistics equivalent to stable distributions
(i.e. Gaussian and L\'{e}vy). The interested reader is referred to \cite{Gumbel67,Castillo88}
for a further discussion of EVT and extreme value distributions. 

A plot of the PDF (\ref{eq:eqn22}) is given in Figure \ref{cap:evtPdf}
for various values of $\Lambda$ and for $\alpha=1.5$. From Figure
\ref{cap:evtPdf} we see that as the value of $\Lambda$ increases,
the PDF of $y_{m}^{*}$ broadens. Importantly, the PDF of $y_{m}^{*}$
(\ref{eq:eqn22}) has an infinite variance and thus has more frequently
occuring extreme values of $y_{m}^{*}$ away from $\bar{y}_{m}^{*}$.
Thus from Figure \ref{cap:evtPdf} and (\ref{eq:eqn23}) we see that
the scatter in the $y_{m}^{*}$ about the average $\bar{y}_{m}^{*}$
increases with $N$ and decreases with $m/N$.

\subsection{Conditioning -- overview}

We now present a method to `condition' data so that the scaling behaviour
(\ref{eq:eqn5pt85}) emerges from the structure functions obtained
for a finite data series. From an operational point of view, that
is, when attempting to determine an (unknown) exponent from a finite
length timeseries, our aim is to recover (\ref{eq:eqn5pt85}) for
as many orders $p$ as feasible. This method involves excluding a
fraction $m/N$ of the largest events from the data set such that
our post-exclusion tails are now sufficiently resolved and populated.
Although there is some literature on the removal of extreme outliers
in data, the first time it was clearly done in the scaling context
was by Veltri \emph{et. al} \cite{Veltri1999,MangeneySalemVeltri2001}.
They calculated structure functions via the use of a Haar wavelet
transform and conditioned their data by separating the wavelet coefficients
into two classes: the majority of coefficients which characterise
the {}``quietly turbulent flow''; and the coefficients which characterise
the rare intermittent events corresponding to coherent structures.
The partition between these two classes was a wavelet coefficient
based upon a multiple $F$ of the square root of the second moment
of the coefficents. The easiest way to view this is by looking at
the more recent works of Chapman \emph{et. al.} \cite{Chapman2005,Hnat2005}
(and refs therein) who employed an equivalent technique but did not
use wavelet transforms to calculate the structure functions. Along
with their solar wind turbulence data, the latter authors also studied
some toy cases of fractional Brownian motion and a L\'{e}vy process of
$\alpha=1.8$. This conditioning can be succinctly written as the
approximation
\begin{eqnarray}
S^{p}(\tau;\pm\infty) & = & \int_{-\infty}^{\infty}\left|y\right|^{p}
P(y,\tau)dy\nonumber \\
 & \rightarrow & S^{C}(\tau;\pm A)\nonumber \\
 & = & \int_{-A}^{A}\left|y\right|^{p}P(y,\tau)dy\ ,
\label{eq:eqn24}
\end{eqnarray}
where $A=Q\sigma(\tau)$, $\sigma(\tau)$ is the standard deviation
and $Q$ is some constant. This corresponds to \emph{clipping} the
wings of the distribution to exclude the very large unresolved events.
Both these studies \cite{Veltri1999,Chapman2005} showed that removing
a relatively few percentage of points is sufficient to regain the
scaling. However, the disadvantage of these schemes is that the measure
used to exclude the extreme events is the standard deviation, $\sigma$,
of the \emph{raw} data which must be calculated \emph{a priori} and
we have already seen in the above analysis that $p>\alpha$ (and thus
$\sigma$) is poorly represented in the unconditioned data. A better
estimate is to condition the data based on the actual extreme events
i.e. by excluding a certain negligible fraction of the data outliers.

A brief mention should be made of the work by Jespersen \emph{et.
al.} \cite{Jespersen1999}. They studied the behaviour of L\'{e}vy flights
in external force fields and used a form of conditioning for obtaining
a good statistical ensemble in the power-law tail range of a L\'{e}vy
process. Their conditioning, however, assumes \emph{a priori} knowledge
of the distribution and its scaling behaviour, and is thus not congruent
to the applications to which this paper aims; this being single finite
size natural timeseries.

To summarise, our procedure will be to:

\begin{enumerate}
\item Choose limits of the integral in (\ref{eq:eqn24}) such that the scaling
(\ref{eq:eqn5pt85}) is recovered -- using a method that does not
require \emph{a priori} knowledge of the PDF $P(y,\tau)$ to specify
those limits.
\item This procedure will exclude the most outlying points ($\lesssim1$\%).
\item These outliers contain some physics of the system. They may or may
not share the scaling (\ref{eq:5pt98}) with the core of the PDF $P(y,\tau)$,
instead showing finite size scaling (exponential roll-off) or other
dynamics. Therefore we will also test the outliers for the property
(\ref{eq:eqn17}).
\end{enumerate}

\subsection{Conditioning -- L\'{e}vy process}

\begin{figure}
\begin{center}a)\includegraphics[%
  width=1.0\columnwidth,
  keepaspectratio]{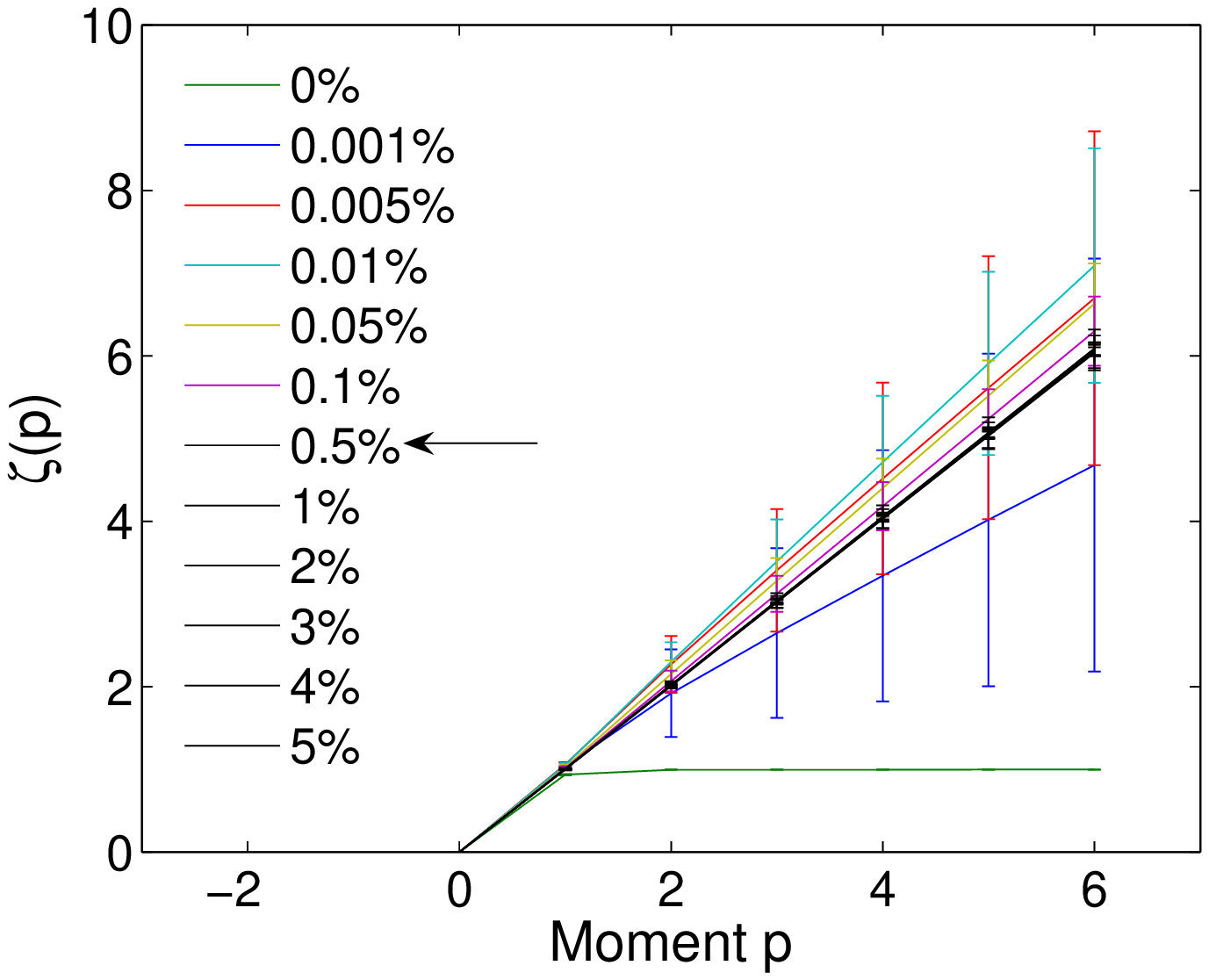}
\end{center}
\begin{center}b)\includegraphics[%
  width=1.0\columnwidth,
  keepaspectratio]{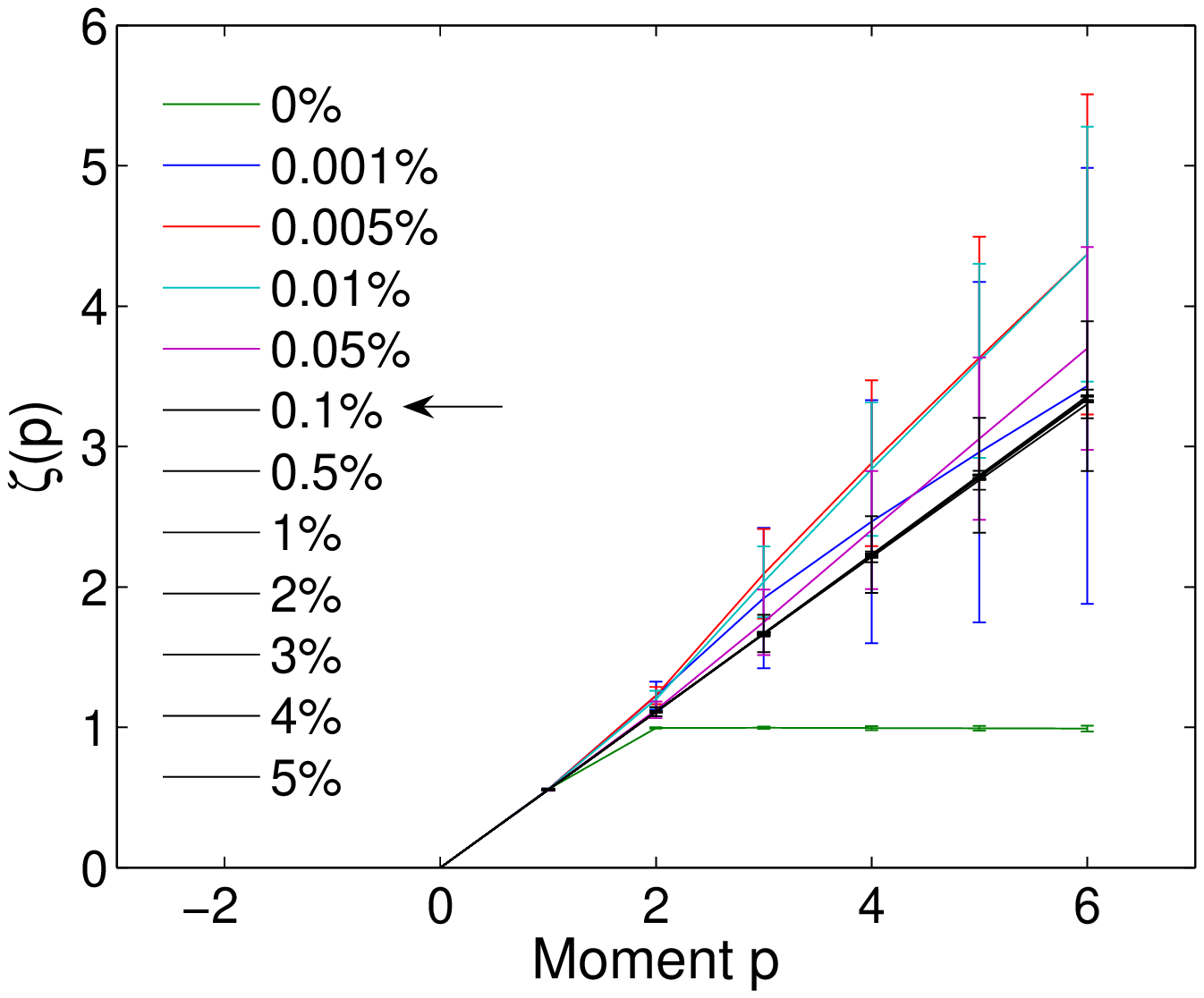}
\end{center}
\caption{\label{cap:zetaplots}Plots showing the exponents $\zeta(p)$ against
moment order $p$ of the generalized structure functions for various
values of the percentage of large events excluded for (a) $\alpha=1.0$
and (b) $\alpha=1.8$. The arrows indicate the percentage beyond which
convergence to the expected behaviour $\zeta(p)=p/\alpha$ is established.
Both plots are for a sample size of $N=10^{6}$.}
\end{figure}

We now test these ideas with a numerically generated L\'{e}vy process.
The increments $y$ of the L\'{e}vy process of index $\alpha$ were generated
by using the following algorithm \cite{SiegertFriedrich2001}
\begin{equation}
y=\frac{\sin(\alpha r)}{(\cos r)^{1/\alpha}}\left(\frac
{\cos\left[(1-\alpha)r\right]}{v}\right)^{(1-\alpha)/\alpha}\ ,
\label{eq:eqn25}
\end{equation}
where $r\in[-\pi/2,\pi/2]$ is a uniformly distributed random variable
and $v$ is an exponentially distributed random variable with unit
mean. Expression (\ref{eq:eqn25}) corresponds to the L\'{e}vy distribution
(\ref{eq:eqn6}) with $\gamma=1$ and $\tau=1$. We generate a sample
of size $N$ and then construct a timeseries by use of a cumulative
sum. This timeseries was then differenced at various $\tau$ as in
(\ref{eq:eqn1}) using an overlapping window; appropriate here since
the data increments are uncorrelated. Structure functions of the increments
$S^{p}(\tau;y_{\pm}^{*}(\tau))$, are then calculated at different
orders $p$ and at different values of $\tau$. These are then plotted
on a $S^{p}$ vs. $\tau$ plot and a linear regression is performed
to obtain the gradients $\zeta(p)$ for each moment order $p$. The
plots of these $\zeta(p)$ vs. $p$ are shown in Figure \ref{cap:zetaplots}
for the two cases $\alpha=1.0$ and $\alpha=1.8$. The error bars
in Figure \ref{cap:zetaplots} were obtained from the difference between
the linear regression of the structure functions for all moment orders
concerned, and the linear regression with the $5^{th}$ and $6^{th}$
moment orders not included.

\begin{figure}
\begin{center}a)\includegraphics[%
  width=1.0\columnwidth,
  keepaspectratio]{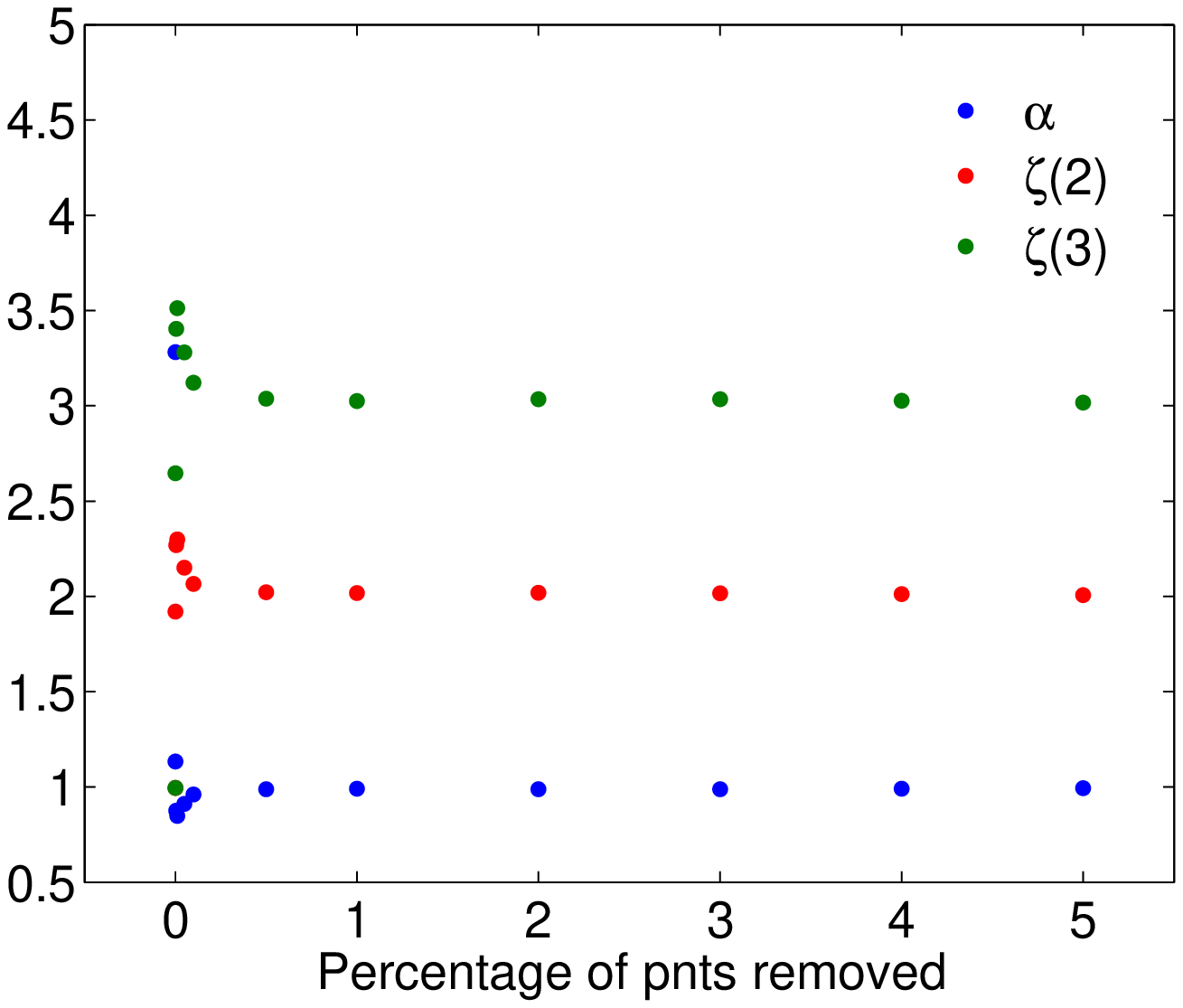}
\end{center}
\begin{center}b)\includegraphics[%
  width=1.0\columnwidth,
  keepaspectratio]{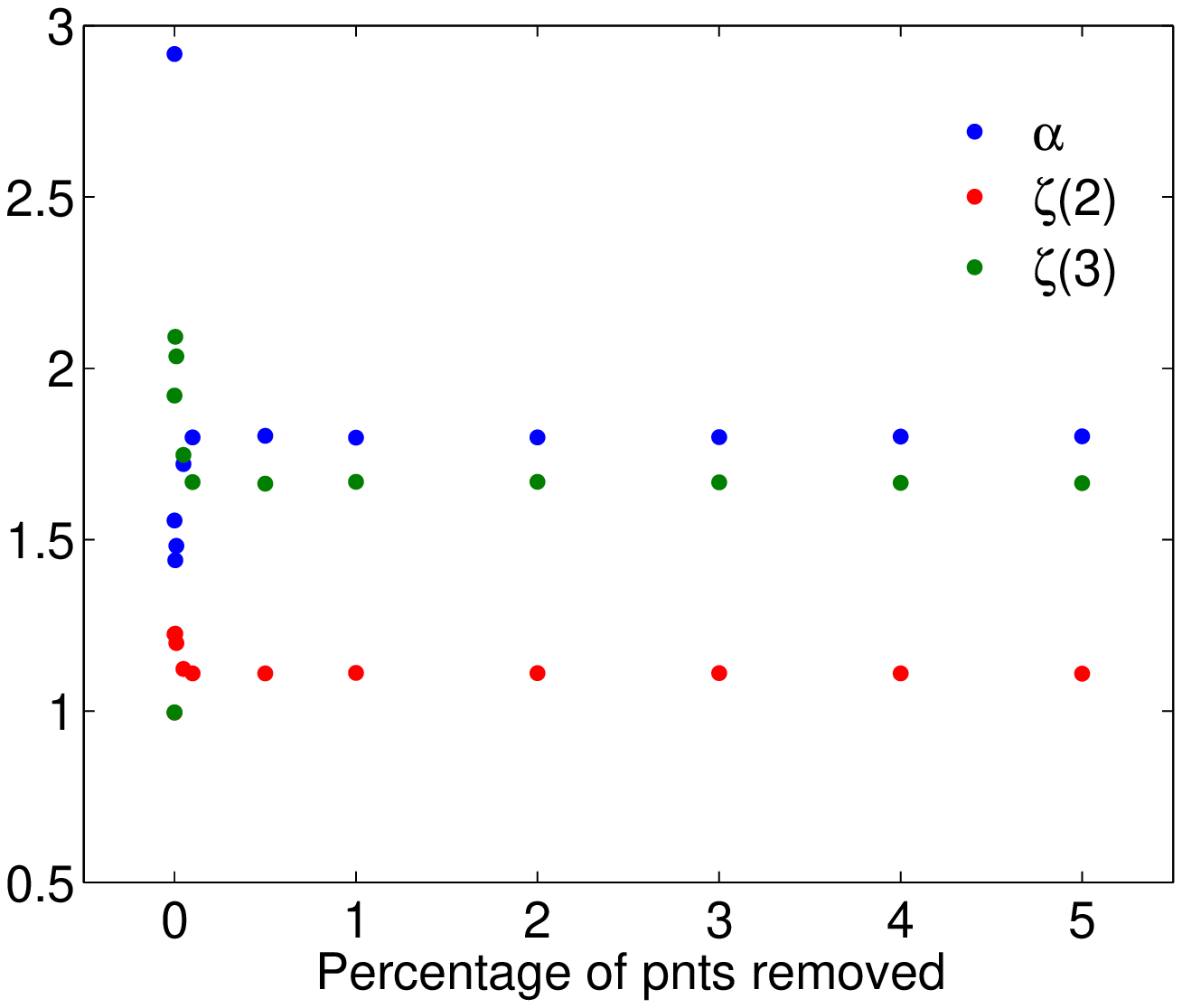}
\end{center}
\caption{\label{cap:alphaConvergence}Plots showing the rapid convergence
of the L\'{e}vy parameter $\alpha$; and the exponents of the $2^{nd}$
and $3^{rd}$ moments $\zeta(2)$ and $\zeta(3)$. The plots in (a)
are for $\alpha=1.0$ and in (b) for $\alpha=1.8$ -- both have $N=10^{6}.$
$\zeta(2)$ and $\zeta(3)$ are the best fit gradients of the $S^{p}$
vs. $\tau$ plots, and $\alpha$ is obtained from the inverse of the
gradient of the $\zeta(p)$ vs. $p$ plot shown in Figure \ref{cap:zetaplots}.}
\end{figure}

In Figure \ref{cap:zetaplots} we see that if no outliers are removed
from the integral for $S^{p}$, the resulting values of $\zeta(p)$
for $p>\alpha$ saturate to unity. Removing a small fraction ($\sim$0.001\%)
of the outliers results in a drastic change in the $\zeta(p)$, again
emphasising the strong effect these points have in the integral for
$S^{p}$. The $\zeta(p)$ converge to the values predicted by (\ref{eq:eqn21})
quite rapidly with $m/N$. The rate of convergence is illustrated
in Figure \ref{cap:alphaConvergence} for the two cases shown in Figure
\ref{cap:zetaplots}. Convergence is achieved at $m/N=0.001$ for
$\alpha=1.8$ and $m/N=0.005$ for $\alpha=1.0$; which correspond to the largest event being 
$y^{*}\simeq18$ and $y^{*}\simeq130$ respectively. These values lie in the 
region given by (\ref{eq:eqn8}), as the asymptotic tail region of the PDF 
is valid for $y\gg\tau^{1/\alpha}=1$ here.

\begin{figure}
\begin{center}a)\includegraphics[%
  width=1.0\columnwidth,
  keepaspectratio]{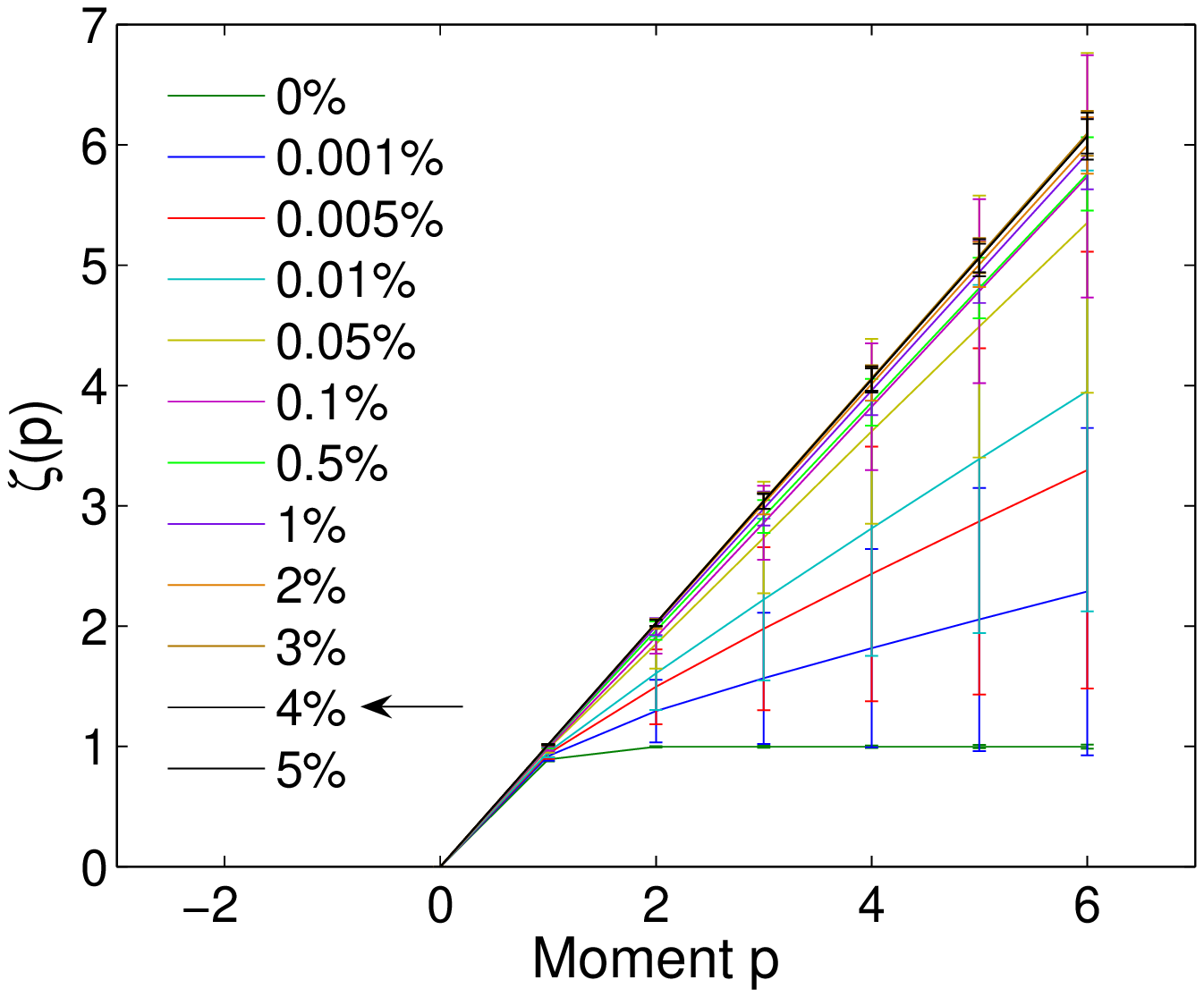}
\end{center}
\begin{center}b)\includegraphics[%
  width=1.0\columnwidth,
  keepaspectratio]{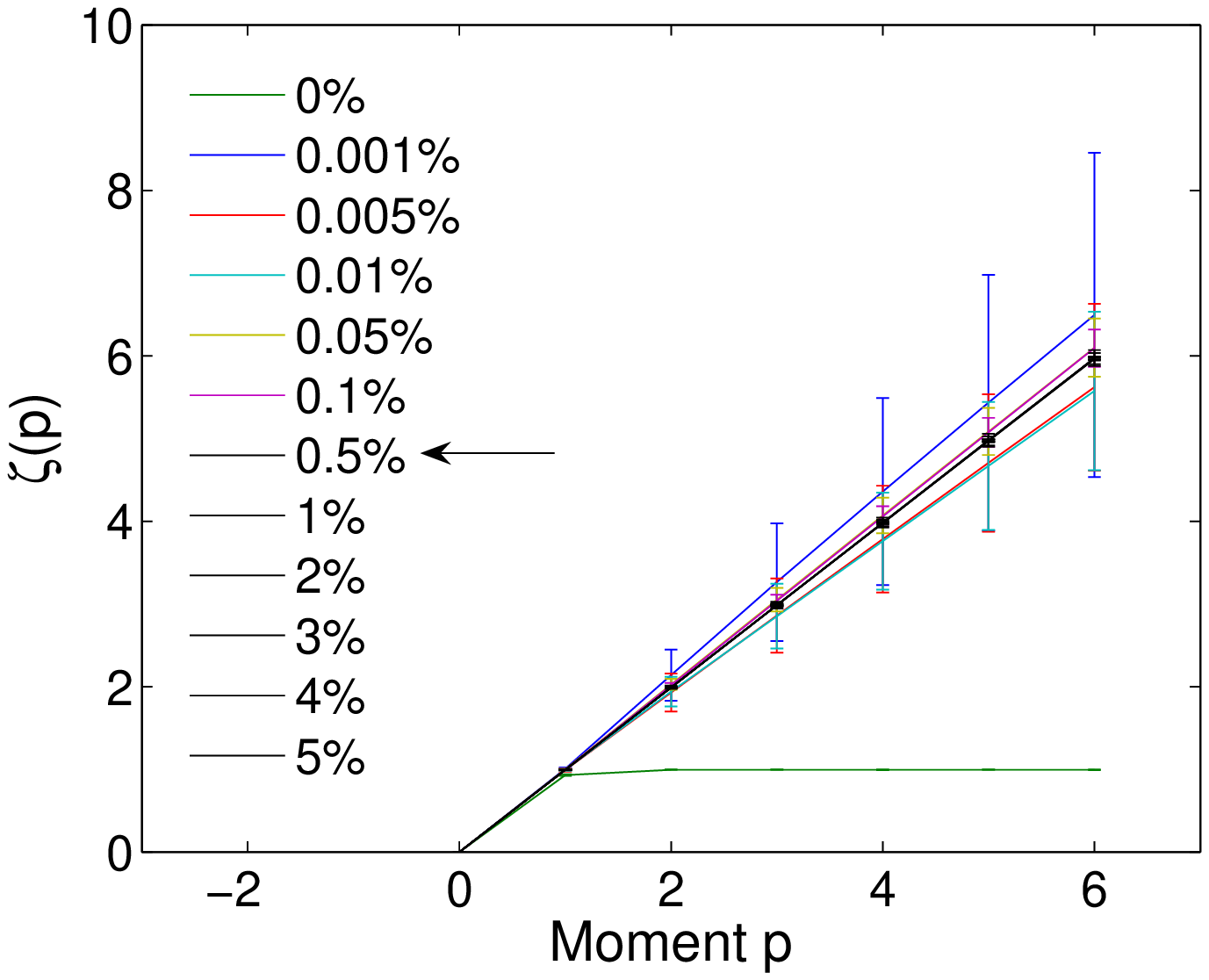}
\end{center}
\caption{\label{cap:gstrfVarN}$\zeta(p)$ vs. $p$ plots for $\alpha=1.0$;
(a) $N=10^{5}$ and (b) $N=5\times10^{6}$.}
\end{figure}

It is also instructive to investigate the effects of variations in
sample size $N$ on the rates of convergence. Figure \ref{cap:gstrfVarN}
illustrates these effects in the form of $\zeta(p)$ vs. $p$ plots
for sizes $N=10^{5}$ and $N=5\times10^{6}$ for a L\'{e}vy process of
index $\alpha=1.0$. Recall that decreasing the sample size would
result in further undersampling and thus poor statistics in the tails
of the PDF. This can be clearly seen in Figure \ref{cap:gstrfVarN}
(a) where we see a slow convergence to the line $\zeta(p)=p/\alpha$
which is achieved after $\sim4$\% of the data is excluded. The converse
of this is shown in Figure \ref{cap:gstrfVarN} (b) where increasing
the sample size by a factor of $20$ results in a very rapid convergence
to scaling which is reached after only $\sim0.5$\% of the data is
excluded. 

\begin{figure}
\begin{center}\includegraphics[%
  width=1.0\columnwidth,
  keepaspectratio]{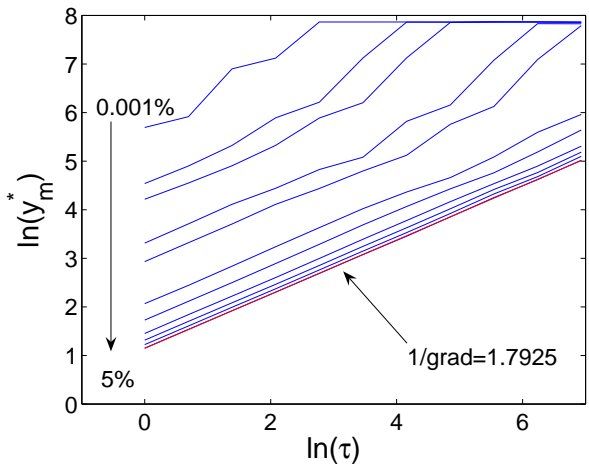}
\end{center}
\caption{\label{cap:xmaxTauN1e6}Log-log plot illustrating the scaling of
the $m^{th}$ largest event $y_{m}^{*}$ with $\tau$ as $m$ is increased;
$\alpha=1.8$ , $N=10^{6}$. For comparison with previous figures
we indicate the \% of points that would be excluded for the particular
$m$.}
\end{figure}

\begin{figure}
\begin{center}\includegraphics[%
  width=1.0\columnwidth,
  keepaspectratio]{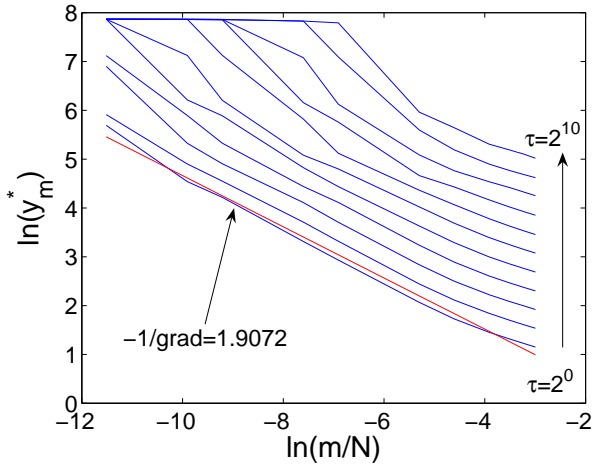}
\end{center}
\caption{\label{cap:xmaxPercN1e6}Log-log plot illustrating the scaling of
the largest event $y_{m}^{*}$ with $m/N$ for various values of $\tau$;
$\alpha=1.8$ , $N=10^{6}$.}
\end{figure}

Lastly we consider the behaviour of the outliers that are removed
by this procedure. As we succesively remove more outliers (increasing
$m$), the behaviour of $y_{m}^{*}$ will more closely correspond
to that of $\bar{y}_{m}^{*}$. This is shown in Figure \ref{cap:xmaxTauN1e6}
where we plot $y_{m}^{*}(\tau)$ for increasing $m/N$. The anticipated
scaling (\ref{eq:eqn17}) appears at a value of $m/N$ corresponding
to a few percent. A more established method for determining the scaling
of outliers is a rank order (or Zipf) plot (see Sornette \cite{Sornette2000});
this is shown in Figure \ref{cap:xmaxPercN1e6} where we plot $y_{m}^{*}(m/N)$
for succesively large values of $\tau$. The scaling with $m/N$ is
again as expected from (\ref{eq:eqn12})--(\ref{eq:eqn16}), and the
rank order plots also highlight scatter of individual realisations
of $y_{m}^{*}$ from the ensemble average. In Figure \ref{cap:xmaxPercN1e6}
this becomes apparent at higher values of $\tau$. As we increase
$\tau$ we require a higher fraction of points to be excluded before
we regain the expected scaling with $m/N$. This breakdown of the
scaling at higher values of $\tau$ follows from equations (\ref{eq:eqn22})
and (\ref{eq:eqn23}). We can see that $\Lambda$ increases with $\tau$
and so the distribution becomes more broad. Consequently this will
require a higher fraction $m/N$ of points to be excluded so that
we may regain the scaling behaviour (\ref{eq:eqn12}). At the largest
$\tau$, Figures \ref{cap:xmaxTauN1e6} and \ref{cap:xmaxPercN1e6}
show a saturation indicative of the difference $y_{m}^{*}$ being
dominated by a single extremal value $x$ of the original timeseries
in (\ref{eq:eqn1}). These plots are also a useful indicator of how
feasable, for a dataset of size $N$, it would be to distinguish a
departure from L\'{e}vy scaling in the tails.


\section{summary and conclusions}

In this paper we have presented a novel technique for `conditioning'
data to deal with anomalous scaling properties that arise due to finite
size effects. We have demonstrated our ideas on a numerically generated
symmetric $\alpha$-stable L\'{e}vy process. We are concerned with the
situation of observations of natural systems, or of experiments, where
the underlying PDF is not known \emph{a priori} and where one inevitably
has a finite length series of data. Hence we have proposed a technique
that does not require \emph{a priori} knowledge of the underlying
process and that has consistency checks.

We have shown that `conditioning' the data by progressively excluding
the outliers, or extremal points, when computing the scaling exponents
from the structure functions, recovers the underlying scaling of a
self-affine process up to large order. For large datasets of a L\'{e}vy
process this corresponds to removing 0.1-1\% of the data. The conditioned
structure functions then provide a straightforward method for determining
the self-affine scaling exponent, in this case the L\'{e}vy index $\alpha$,
directly from the slope of a plot of the exponents versus moment order.

This method offers two consistency checks. The first of these is that
for a self-affine process, as we progressively remove more outliers
we expect that the exponents obtained from the structure functions
should converge on values which then do not vary. Practically speaking,
one would plot the exponents as a function of the location of the
last outlier excluded and expect a plateau that extended deep into
the tail of the PDF. A second check is obtained by examining the scaling
properties of these discarded outliers. 

Importantly, the above analysis assumes that we have some relatively
good statistics -- in practice the high variability of the L\'{e}vy process
due to the fat tails will always result in some lone extreme points
with a finite probability of occurence, resulting in anomalous scaling
exponents. This implies that we \emph{always} need some way of cleaning
or conditioning the data to recover the scaling behaviour. These lone
points can have a drastic effect since in a L\'{e}vy-like process the
largest value of a set of increments of a timeseries can be of the
order of the total sum \cite{BardouBouchaud02,Sornette2000}. Coupled
with this we have that the tails of a distribution are described by
the higher order moments (structure functions here). If the statistics
of the tail are not well resolved then these moments will also give
anomalous values of $\zeta(p)$.

In principle, this approach may be extended to the case of multi-affine
timeseries and this will be the subject of further work.

\begin{acknowledgments}
The authors would like to thank N. Watkins and G. Rowlands for helpful
discussions and suggestions. KK acknowledges the financial support
of the Particle Physics and Astronomy Research Council.

\bibliographystyle{revtex}

\end{acknowledgments}

\end{document}